\documentclass[twocolumn,aps,showpacs,epsfig]{revtex4}

\usepackage{amsmath}
\usepackage[dvips]{graphics}
\usepackage{latexsym}
\usepackage{subfigure}

\DeclareFontFamily{OT1}{rsfs}{}
\DeclareFontShape{OT1}{rsfs}{m}{n}{ <-7> rsfs5 <7-10> rsfs7 <10->
rsfs10}{} \DeclareMathAlphabet{\mycal}{OT1}{rsfs}{m}{n}
\def\scri{{\mycal I}}

\begin{document}
\newcommand{\bea}{\begin{eqnarray*}}
\newcommand{\eea}{\end{eqnarray*}}
\newcommand{\bean}{\begin{eqnarray}}
\newcommand{\eean}{\end{eqnarray}}
\newcommand{\eqs}[1]{Eqs. (\ref{#1})} 
\newcommand{\eq}[1]{Eq. (\ref{#1})} 
\newcommand{\meq}[1]{(\ref{#1})} 
\newcommand{\fig}[1]{Fig. \ref{#1}}

\newcommand{\tri}{\delta}
\newcommand{\grad}{\nabla}
\newcommand{\pa}{\partial}
\newcommand{\pf}[2]{\frac{\pa #1}{\pa #2}}
\newcommand{\cla}{{\cal A}}
\newcommand{\aqt}{\frac{1}{4}\theta}

\newcommand{\om}{\omega}
\newcommand{\omo}{\omega_0}
\newcommand{\ep}{\epsilon}
\newcommand{\nonu}{\nonumber}
\newcommand{\scrip}{\scri^{+}}
\newcommand{\hp}{{\cal H^+}}
\newcommand{\tm}{\tilde M} 
\newcommand{\ts}{\frac{\sqrt 3}{2}}
\newcommand\tabcaption{\def\@captype{table}\caption}

\title{The Tolman-Bondi--Vaidya Spacetime: 
matching timelike dust to null dust}

\author{Sijie Gao}
\email{sijie@fisica.ist.utl.pt}
\author{Jos\'e P. S. Lemos}
\email{lemos@fisica.ist.utl.pt}
\affiliation{
Centro Multidisciplinar de Astrof\'{\i}sica - CENTRA,\\
Departamento de F\'{\i}sica, Instituto Superior T\'ecnico,\\
Universidade T\'ecnica de Lisboa,\\
Av. Rovisco Pais 1, 1049-001 Lisboa, Portugal}

\begin{abstract}
The Tolman-Bondi and Vaidya solutions are two solutions to Einstein
equations which describe dust particles and null fluid,
respectively. We show that it is possible to match the two solutions
in one single spacetime, the Tolman-Bondi--Vaidya spacetime.  The new
spacetime is divided by a null surface with Tolman-Bondi dust on one
side and Vaidya fluid on the other side. The differentiability of the
spacetime is discussed. By constructing a specific solution, we show
that the metric across the null surface can be at least $C^1$ and the
stress-energy tensor is continuous.

\end{abstract}


\maketitle

\section{Introduction}
The Tolman-Bondi solution describes the inhomogeneous gravitational
collapse of spherical dust. In comoving coordinates, the metric is
given by \cite{tolman, lemosA} 
\bean
ds^2=-dt^2+\frac{r'^2}{1+E(R)}dR^2+r^2d\Omega^2, \label{tb}
\eean
where $d\Omega^2$ is the metric of the unit two-sphere,  $E(R)$ is an
arbitrary function of the comoving radial coordinate $R$, 
and the prime denotes the partial 
derivative with respect to $R$. The function $r(t,R)$ is a solution of 
\bean
\dot r^2=\frac{2M(R)}{r}+E(R) , \label{rd}
\eean
where the overdot denotes the partial derivative with respect to $t$
and $M(R)$ is the effective gravitational mass within $R$. In
contrast, the incoming Vaidya metric describing collapsing null dust
is given by \cite{vaidya, hiscock}
\bean
ds^2=-\left(1-\frac{2M(v)}{r}\right)dv^2+2dvdr+r^2d\Omega^2\,, \label{vai}
\eean
where $v$ is an advanced time and the $v={\rm constant}$ line gives
the trajectory of the null fluid.  Both the Tolman and the Vaidya
spacetimes provide important examples for the study of naked
singularities formed in gravitational collapse. The intrinsic relation
between the two solutions had not been investigated for many
years. Lemos first showed in \cite{lemos}, that the Vaidya metric
\meq{vai} can be obtained from the Tolman-Bondi metric \meq{tb} by
taking the limit $E\rightarrow \infty$. Therefore, the two metrics
belong to one family and their naked singularities are of the same
nature. Hellaby \cite{hellaby} pointed out that Lemos' derivation
requires that $E$ be a constant through the spacetime. As a
generalization, Hellaby showed that this assumption can be dropped and
the same Vaidya limit can be achieved from the Tolman-Bondi solutions
by using a different coordinate transformation.

Despite the connection
discovered above, the two solutions have been treated in two separate
spacetimes which are connected only mathematically by taking the limit
on $E$. Is it possible to put the two solutions in one single
spacetime such that the Tolman-Bondi dust can approach the Vaidya
fluid as a physical limit? Hellaby's work has provided some
necessary mathematical framework to solve this problem. Hellaby
makes no assumptions on the function $E(R)$ in the Tolman-Bondi
model. This is an important point for our generalization. In \cite{hellaby},
the Vaidya limit is obtained by taking the limit
$E(R)\rightarrow \infty$ everywhere. 

Our approach to matching the two
solutions in one spacetime is simple. Instead of
requiring  $E(R)\rightarrow \infty$ 
everywhere, we require $E(R)$ to diverge only at one hypersurface
$R=R_0$. For $R<R_0$, $E(R)$ is finite and it corresponds to a
Tolman-Bondi solution. For $R>R_0$, the spacetime transfers into the
Vaidya null fluid. In this new spacetime, dust particles which follow
timelike geodesics and photons which follow null geodesics
coexist and a phase transition occurs across the $R=R_0$
hypersurface. By imposing some general conditions, the spacetime is
continuous ($C^0$) by construction. 
An important issue to be discussed is the smoothness
around the limit surface $R=R_0$, i.e., how well one can match
the two solutions together. We show from a specific example, that the
metric across the limit surface can be at least $C^1$. We also
show that the stress-energy tensor and the Kretschmann scalar are both
continuous in the spacetime.

\section{Construction of the spacetime} \label{after}

In this section, we shall construct a spherically symmetric spacetime
which combines the Tolman-Bondi dust and the Vaidya null dust. To be
definitive, the Tolman-Bondi dust will be configured inside the Vaidya
dust. However, all of our arguments and results apply to the reversed
configuration.  We start with the Tolman-Bondi metric \meq{tb}. It is
evident from this metric that each comoving observer follows a
timelike worldline with $R={\rm constant}$. To approach the Vaidya
limit, we shall let $E(R)$ diverge at some fixed $R=R_0>0$ such that
the spacetime consists of the Tolman-Bondi dust for $R<R_0$ and the
Vaidya dust for $R\geq R_0$. If this can be done, the $R=R_0$ surface
will become a limit null surface for Tolman-Bondi observers, which
indicates that the original Tolman-Bondi coordinates in \eq{tb} will
break down at $R=R_0$. This is reminiscent of the Schwarzschild
solution where the original Schwarzschild coordinates break down when
one approaches the event horizon, and the introduction of the Kruskal
extension is necessary to eliminate the apparent coordinate
singularity. Now we face a similar situation and we shall introduce a
new set of coordinates to extend the spacetime from the Tolman-Bondi
interior to the Vaidya exterior, to yield a Tolman-Bondi--Vaidya
spacetime.  Following \cite{hellaby}, we rewrite \eq{tb}, 
eliminating $dt$ through
\bean
dr=\dot r dt+r'dR\;\rightarrow \;dt=\frac{-r'dR+dr}{\dot r} \label{dR}
\eean
and then using \eq{rd}, to obtain 
\bean
ds^2 &=&
-\left(1-\frac{2M}{r}\right)
\frac{r'^2}{\left[1+E(R)\right]\dot r^2}dR^2 
+2\frac{r'}{\dot  r^2}\,dR\,dr \nonumber \\
&&-\frac{1}{\dot r^2}dr^2 
+r^2d\Omega^2\,. \label{dsr}
\eean
Now the metric is expressed in terms of $(R,r,\theta,\phi)$. Since $r$
is the areal radius, it should be kept as a good coordinate in the
extended spacetime. We note, in passing, that Hellaby further
introduced a new coordinate $v$ defined by $v=\int_0^R \frac{a'}{\sqrt
E}dR-\frac{r}{2E}\,,$ where $a$ is some function of $R$, such that the
metric is finally expressed in coordinates
$(v,r,\theta,\phi)$. However, the components of the metric involve
functions like $E(R)$ and $M(R)$ that are implicit functions of
$(v,r)$. Consequently, it will be difficult to study the
differentiability of the metric with Hellaby's coordinate
system. Indeed we find that introducing the new coordinate $v$ is not
necessary here. Under certain conditions, coordinates
$(R,r,\theta,\phi)$ are good enough for our purposes to describe the
spacetime. To proceed, we first give some limiting forms of the
components of the metric \meq{dsr} which has been derived by Hellaby
in \cite{hellaby}.

The exact solution for $r'$ in a collapsing model is \footnote{In
Eq.(10) of \cite{hellaby}, the sign before the bracket ``[''is
positive. We believe it is a misprint and have corrected it below.}  
\bean
r'=\left(\frac{M'}{M}-\frac{E'}{E}\right)r-\left[a'-\left(\frac{M'}{M}-
\frac{3E'}{2E}\right)(a-t)\right]\dot r, \label{exrp}  
\eean
where $a=a(R)$ is another arbitrary function. We shall show in section
\ref{sub2}, that  $t$ can be written as an explicit function of
$(R,r)$. Thus, each component in the metric of \eq{dsr} depends on
$(R,r,\theta,\phi)$. Hellaby required that $M$ and $r$ be
finite while $E$ goes to infinity everywhere. Here we adopt a
similar requirement with only a small modification, albeit important: we
require  $M$ 
and $r$ remain finite everywhere and $E(R)$ is divergent for $R\geq
R_0$. Then we have the following limiting forms at $R=R_0$ \cite{hellaby}:
\bean
r&\rightarrow& \sqrt E (a-t) \label{lr} \nonumber\\
\dot r&\rightarrow&-\sqrt E \label{lrd}\\
r'&\rightarrow& \frac{rE'}{2E}+a'\sqrt E.  \nonumber\label{lrp}
\eean
By simple substitution, we find the following limits for the components
of \eq{dsr} (see \cite{hellaby})
\bean
\frac{1}{\dot r^2}&\rightarrow& \frac{1}{E}, \nonumber\\
\frac{r'}{\dot r^2}&\rightarrow& \left(\frac{a'}{\sqrt
  E}+\frac{rE'}{2E^2}\right),\\
\frac{r'^2}{(1+E)\,\dot r^2}&\rightarrow&
 \left(\frac{E}{1+E}\right)\left(\frac{a'}{\sqrt 
  E}+\frac{rE'}{2E^2}\right)^2\,.  \nonumber
\eean
If we further impose restrictions such that 
\bean
\frac{a'}{\sqrt E}+\frac{rE'}{2E^2}\rightarrow 1 \quad {\rm at}
\quad R=R_0\,,\label{apo}
\eean
and identical to $1$ for $R>R_0$, one sees immediately, by
comparing the components of the original Tolman-Bondi metric \meq{dsr}
and those of the Vaidya 
metric \meq{vai} and identifying $v$ with $R$, that the metric of
\eq{dsr} becomes exactly the Vaidya metric for $R>R_0$. 
In summary, our spacetime is constructed in the following way: The
coordinates are $(R,r,\theta,\phi)$. For $R<R_0$, the spacetime is
described by any Tolman-Bondi dust in the form of
\meq{dsr} which satisfies the limiting condition \meq{apo} and
$E\rightarrow \infty$ at $R=R_0$. For $R>R_0$, the spacetime is
described by the Vaidya null dust \meq{vai} with $v$ relabeled by
$R$. In this case, we can choose a smooth mass function
$M(R)$ for the entire spacetime. With this choice, the spacetime metric is
continuous (or $C^0$) by construction. For future reference, we write the
complete metric as follows
\begin{widetext}
\bean
ds^2=\left\{ \begin{array}{ll}
-\left(1-\frac{2M(R)}{r}\right) \frac{r'^2}{[1+E(R)]\,\dot r^2}dR^2
+2\frac{r'}{\dot r^2}\,dR\,dr
-\frac{1}{\dot r^2}dr^2 +r^2d\Omega^2\,, & R<R_0\,, \\
-\left(1-\frac{2M(R)}{r}\right)dR^2+2\,dR\,dr+r^2d\Omega^2\,,  
& R>R_0 \,.\label{big}
\end{array}\right. 
\eean
\end{widetext}
We see that  the comoving coordinate $R$ of the Tolman-Bondi dust
automatically becomes the comoving coordinate of the Vaidya null
dust. This shows the unification of the two solutions expressed in
this coordinate system. It is also easy to check that the $RR$ component
of the inverse metric, $g^{RR}$, approaches zero as $R\rightarrow
R_0$, which confirms that 
the hypersurface $R=R_0$ is a null hypersurface. In the next section, we
shall show through a concrete example that the metric at $R=R_0$ can be
at least $C^1$.

\section{An example}
To demonstrate how well the two metrics can be matched, we provide an
explicit example based on the prescription given in the previous
section. 

\subsection{Choosing the arbitrary functions} \label{sub1}
Since our interest is matching the Tolman-Bondi and
Vaidya solutions, we shall restrict our attention to the neighborhood
of the limit null hypersurface, which is denoted by
$\Sigma$. Since at $R=R_0$ the function $E(R)$ goes to infinity, 
it is convenient to discuss the smoothness of $E(R)$ at $R=R_0$ 
through an auxiliary function $h(R)$ given by
\bean
h(R)=\frac{1}{E(R)}.   \label{hR}
\eean 
According to the discussion in the previous section, $h(R)$ should
approach zero at $\Sigma$ from the Tolman-Bondi interior ($R<R_0$) and
be identical zero in the Vaidya exterior. Hence, we choose $h(R)$ as 
\bean
h(R)=\left\{ \begin{array}{ll}
 (R_0-R)^3 & R<R_0 \\
0& R\geq R_0.    \label{dhR}
\end{array}\right. 
\eean
To satisfy the limit \meq{apo}, we define $a(R)$ in the Tolman-Bondi
interior as 
\bean
a(R)=\frac{2}{\sqrt{R_0-R}}. \label{aR}
\eean
In the following two subsections, we shall investigate the
differentiability of the metric associated with the model spacetime and the
continuity of the Einstein tensor across the null surface $\Sigma$.

\subsection{Differentiability of the metric} \label{sub2}

Our task in this subsection is to show that the metric is $C^1$ on the
null surface $\Sigma$. We shall show that the components of the metric
is differentiable on $\Sigma$.

We begin with the $RR$ component in \eq{big}, denoting it on the
Tolman-Bondi side, $R<R_0$, by $g^-_{RR}$, and on the Vaidya side,
$R>R_0$, by $g^+_{RR}$.  Combining \eqs{rd},\meq{dsr}, \meq{hR},
\meq{dhR} and \meq{aR}, one may write  $g^-_{RR}$ as a function of $(R,r)$
\begin{widetext}
\bean
g^-_{RR}&=&\frac{-1+\frac{2M(R)}{r}}{\left(1+\frac{1}{(R_0-R)^3}\right)
\left(\frac{1}{(R_0-R)^3}+\frac{2M(R)}{r}\right)}\times 
\left[r\left(\frac{3}{R-R_0}+\frac{M'(R)}{M(R)}\right)+
\sqrt{\frac{1}{(R_0-R)^3}+\frac{2M(R)}{r}}\right.  \nonumber \\
&&\left. \left(\frac{1}{(R_0-R)^{3/2}}-
\left(\frac{2}{\sqrt{R_0-R}}-t(R,r)\right)\left(\frac{9}{2R-2R_0}+
\frac{M'(R)}{M(R)}\right)\right)\right]^2 \,,  \label{lgRR}    
\eean 
\end{widetext}
where we have expressed $t$ formally as a function of $(R,r)$. To find
the explicit form of $t(R,r)$, we consider the hyperbolic solution for
the collapsing model \cite{hellaby}
\bean
r&=&\frac{M}{E}(\cosh\eta-1)\,, \label{sone} \\
(\sinh\eta-\eta)&=&\frac{E^{3/2}(a-t)}{M}\,. \label{stwo}
\eean
Solving \eq{sone} for $\eta$ gives 
\bean
\eta=\cosh^{-1}\left(1+\frac{Er}{M}\right)\,,
\eean
where $\cosh^{-1}$ is the inverse function of $\cosh$ and is required
to be positive. By symmetry, there is also a negative solution for
$\eta$. But one can easily check that, as $E\rightarrow \infty$, the
negative $\eta$ leads to $r\rightarrow \sqrt{f}(t-a)$, which is not a
collapsing solution. Therefore $\eta$ is large and positive in the
region of interest, which implies that $\sinh\eta$ is positive
too. Thus, from \eqs{sone} and \meq{stwo}, we find
\bean
t=a-\frac{\sqrt{E r(2M+E r)}-M\cosh^{-1}(1+Er/M)}{E^{3/2}}\,. \label{trr} 
\eean
It is obvious that $t$ diverges at $R=R_0$ for the functions chosen in 
section \ref{sub1}. This confirms the breakdown of the original
coordinates, as we have mentioned at the beginning of section
\ref{after}. Now we have $g^-_{RR}$ as an explicit function of
$(R,r)$. Its derivatives can then be calculated straightforwardly. To
obtain the limits of these derivatives at $R=R_0$, we treat $R_0-R$ as a
small quantity and compute the power series expansion about
$R=R_0$. This treatment 
guarantees the accuracy of the final results because no approximation
has to be made in the intermediate steps. With the help
of Mathematica, the results can be obtained almost instantly. We find that at
$R=R_0$,  
\bean
\frac{\partial g^-_{RR}}{\partial r}&=&\frac{\partial
g^+_{RR}}{\partial r}=-\frac{2M(R_0)}{r^2}\,,  \nonumber \\
\frac{\partial
g^-_{RR}}{\partial R}&=&\frac{\partial 
g^+_{RR}}{\partial R}=\frac{2M'(R_0)}{r}\,.
\eean
Therefore, the $RR$ component
of the metric is differentiable (or $C^1$) at $\Sigma$. 

With similar notations, we repeat the procedure above for $g^-_{Rr}$
and $g^-_{rr}$ and find that their derivatives with respect to $R$ and
$r$ vanish as $R\rightarrow R_0$, agreeing with their Vaidya
counterparts (note that  $g^+_{Rr}$ and  $g^+_{rr}$ are both
constants). Therefore, we have shown that the spacetime metric is
$C^1$ across the null hypersurface $\Sigma$.

However, it is worth mention, as a remark, that there are examples
where the metric is not always $C^1$ on $\Sigma$ when the conditions
in section \ref{after} are satisfied. Here is a counterexample. We
choose, in the $R<R_0$ region, $h(R)=(R_0-R)^2$ and
$a(R)=-\ln(R_0-R)$. 
It is easy to check that the limiting condition \meq{apo} is satisfied
for this choice. By repeating the similar calculation we carried out
for the first example, we find that ${\partial g^-_{Rr}}/{\partial
R}\rightarrow -r$ when $R\rightarrow R_0$, which is not equal to the
${g^+_{Rr}}/{\partial R}\rightarrow 0$ limit except for the $r=0$
singularity. Thus, this example shows that the metric is not
differentiable (or $C^1$) on $\Sigma$. There is a qualitative
explanation for why the first example works while the second one does
not. Since $h(R)$ vanishes identically for $R>R_0$, we see $h(R)$
defined by \eq{dhR} is $C^2$ at $R=R_0$ while $h(R)$ defined above is
just $C^1$. Roughly speaking, a better behaved function corresponds to
a better behaved metric.

\subsection{Continuity of the stress-energy tensor and the \break
Kretschmann scalar}
We have shown that the spacetime metric defined at the beginning of
section \ref{after} is at least $C^1$. In this subsection, we shall
show that the spacetime performs even better than  $C^1$. One
important quantity associated with the second-order derivatives of the
metric is the stress-energy tensor $T_{ab}$, which can be derived from
Einstein's equation
\bean
G_{ab}\equiv R_{ab}-\frac{1}{2}Rg_{ab}=8\pi T_{ab}. \label{ein}
\eean
We now calculate $G_{ab}$ associated with the metric defined by
\eq{big} and functions chosen in section \ref{sub1}. From Einstein's
equation, it is straightforward to check that 
the only nonzero component of $G_{ab}$ for the Vaidya metric is 
\bean
G_{RR}=\frac{2M'(R)}{r^2}. \label{gvai}
\eean
We need to show that $G_{ab}$ associated with the Tolman-Bondi metric
agrees with this limit at $R=R_0$. The calculation follows the same
strategy as in subsection \ref{sub2}. The limits are taken only in the
final step by means of power expansion. We finally find that the 
Tolman-Bondi stress-energy tensor approaches the Vaidya limit at
$R=R_0$, i.e., all components of $G_{ab}$ vanish at $R=R_0$ except
$G_{RR}\rightarrow {2M'(R_0)}/\,{r^2}$. 

Another quantity related to the spacetime curvature is the Kretschmann
scalar $K=R^{abcd}R_{abcd}$, which is given in \cite{hellaby} for the
Tolman-Bondi metric 
\bean
K^T=\frac{48 M^2}{r^6}-\frac{32 M
  M'}{r^5r'}+\frac{12M'^2}{r^4r'^2}\,\label{kt}
\eean
and for the Vaidya metric
\bean
K^V=\frac{48 M^2}{r^6}\,. \label{kv}
\eean
Since $r'\rightarrow \infty$ in our model spacetime, we see that the
Kretschmann scalar is  continuous across $\Sigma$. 

\section{Concluding remarks}
We have constructed, by generalizing Lemos and 
Hellaby's methods \cite{lemos,hellaby}, a
spacetime where the Tolman-Bondi dust and the Vaidya null fluid
coexist in harmony. The spacetime is described by a uniform
coordinate system which is comoving with both the Tolman-Bondi and
Vaidya observers. We have shown, from an explicit example, that the
spacetime metric is at least $C^1$ across the null surface
$\Sigma$. Moreover, we
have shown that the stress-energy tensor and the Kretschmann are continuous on
$\Sigma$. This property fails for some other spacetimes. For instance,
when matching the Friedmann interior to the Schwarzschild exterior
\cite{mtw}, the energy density across the matching surface is
obviously discontinuous since the Friedmann interior has a uniform
distribution of density and the Schwarzschild exterior has a vanishing
density. Although the Tolman-Bondi and Vaidya models are two
well-known solutions to Einstein's equations, the match of the two models
is a new issue. We have studied the possibility of matching them in one
spacetime. It remains unknown to us how this match can be realized in a
physical process, or what physical mechanism can make it happen.

\begin{acknowledgments}

This work was partially funded by Funda\c c\~ao para a Ci\^encia e
Tecnologia (FCT) through project PDCT/FP/50202/2003. 
SG acknowledges financial support from FCT
through grant SFRH/BPD/10078/2002. JPSL thanks Observat\'orio Nacional
do Rio de Janeiro for hospitality. 

\end{acknowledgments}

\end{document}